 \documentclass[apj]{emulateapj}

\shorttitle{Infrared Imaging of Herschel~36~SE}
\shortauthors{Goto et al.} 

\begin{document}

\title{High-Resolution Infrared Imaging of Herschel~36 SE: \\ 
A Showcase for the Influence of Massive Stars in Cluster Environments
\altaffilmark{1,2}}

\author{M. Goto,\altaffilmark{3} B. Stecklum,\altaffilmark{4}
H. Linz,\altaffilmark{3,4} M. Feldt,\altaffilmark{3}
Th. Henning,\altaffilmark{3} I. Pascucci,\altaffilmark{3,5}, \protect\and
T. Usuda\altaffilmark{6}} 

\email{mgoto@mpia-hd.mpg.de}

\altaffiltext{1}{Based on data collected in the course of NACO and
                 MIDI guaranteed time observations (71.C-0143(A) and
                 73.C-0175(A)) at the VLT on Cerro Paranal (Chile),
                 which is operated by the European Southern
                 Observatory (ESO).}

\altaffiltext{2}{Based on data collected at Subaru Telescope, which is
                 operated by the National Astronomical Observatory of
                 Japan.}

\altaffiltext{3}{Max Planck Institute for Astronomy, 
                 K\"onigstuhl 17, D-69117 Heidelberg, Germany.}
     
\altaffiltext{4}{Th\"uringer Landessternwarte Tautenburg, Sternwarte 5,
                 D-07778 Tautenburg, Germany.}

\altaffiltext{5}{The University of Arizona, Department of
                 Astronomy/Steward Observatory, 993 N Cherry
                 Ave. Tucson, AZ 85721-0065.}

\altaffiltext{6}{Subaru Telescope, 650, North A`ohoku Place, Hilo, HI
                 96720.}

\begin{abstract}

We present high-resolution infrared imaging of the massive
star-forming region around the O-star Herschel~36. Special emphasis is
given to a compact infrared source at 0\farcs25 southeast of the
star. The infrared source, hereafter Her~36~SE, is extended in the
broad--band images, but features spatially unresolved Br$\,\gamma$
line emission. The line-emission source coincides in position with the
previous {\it HST} detections in H$\,\alpha$ and the 2~cm radio
continuum emission detected by {\it VLA} interferometry. We propose
that the infrared source Her~36~SE harbors an early B-type star,
deeply embedded in a dusty cloud. The fan shape of the cloud with
Herschel~36 at its apex, though, manifests direct and ongoing
destructive influence of the O7V star on Her~36~SE.

\end{abstract}

 \keywords{circumstellar matter --- dust, extinction --- planetary
   systems: protoplanetary disks --- early-type --- stars: formation
   --- stars: individual (Herschel~36, G5.97$-$1.17)}

\section{Introduction}

Massive stars are the primary source of radiation, kinetic energy, and
chemical enrichment in the interstellar medium, playing a pivotal role
in galactic evolution. Because of their remote locations, our
understanding of their formation has been limited by the lack of high
resolution techniques. This challenge has been undertaken by adaptive
optics systems at large-aperture telescopes. The present work is part
of a coordinated effort to understand the formation of high-mass stars
by using state-of-the-art instruments available at the {\it VLT} and
other telescopes \citep{fel99,hen01,fel03,gra04,pas04,pug04,lin05,apa05}.

Herschel~36 is located in a high-mass star-forming region at a
distance of 1.8~kpc from us (van den Anker et al. 1997; but see also
Arias et al. 2006) near the center of M~8. The bright central part of
M~8 is called the Hourglass. The Hourglass is a cavity of ionized gas
seen through the gaps between the foreground obscuration
\citep{woo86}. Herschel~36, an O7V star \citep{woo61}, is responsible
for the ionization of the gas in the cavity. The inferred dynamical
age of the ionized gas and, therefore, the age of the Hourglass and
Herschel~36, is as small as $\sim$5$\times$10$^4$ yrs \citep{cha97}.

In the present study a special focus is placed on the infrared source
found at a distance of 0\farcs25 southeast of Herschel~36. The
extended source called hereafter Her~36~SE was first recognized by
\citet{ste95} by means of lunar occultation measurements. Herschel~36
had long been known as a peculiar early-type star with substantial
mid-infrared excess \citep{woo73,all86}. It is only after Her~36~SE
was spatially resolved that we know that this object is actually the
source responsible for the excess infrared emission. After the
discovery of \citet{ste95}, the possible identity of Her~36~SE has
been discussed, including an externally ionized protoplanetary disk,
or, a proplyd, an obscured embedded source, and a leftover
circumstellar disk of Herschel~36; however, no solid conclusion was
reached.

In the next section, the observations at the {\it VLT} and
supplemental spectroscopy at the Subaru Telescope are described.  The
direct consequences of the observations are summarized in \S 3. In \S
4 we will further discuss the possible nature of Her~36~SE as a deeply
embedded early-type star under the violent influence of the nearby
O-star Herschel~36.

\figurenum{1}
\begin{figure}[tbh]
\includegraphics[angle=0,width=0.49\textwidth]{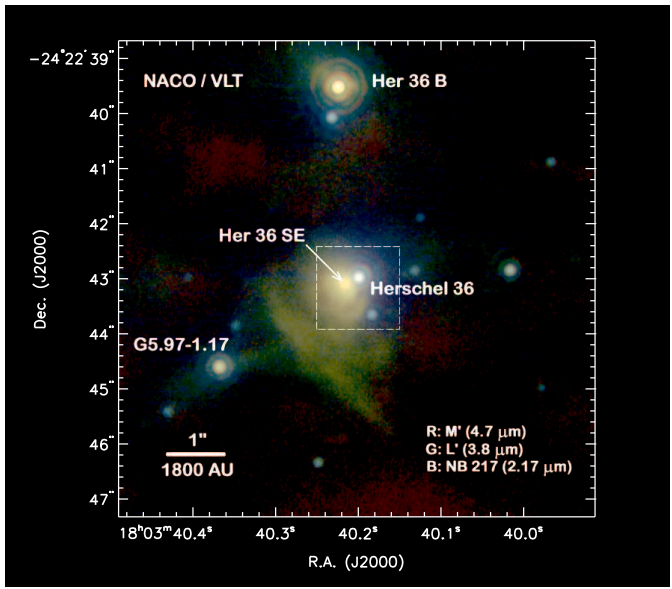}
\caption{Color composite image of Herschel~36. $M'$ is color coded in
  red, $L'$ in green, and Br$\,\gamma$ in blue before continuum
  subtraction. Her~36~SE is the red extended emission at 0\farcs25
  southeast of Herschel~36.  Her~36~B is an infrared star at 3\farcs6
  north of Herschel~36. The coordinates of Herschel~36 are R.A.~(J2000)
  18$^{\rm h}$03$^{\rm m}$40$.\!\!^{\rm s}$20, Dec (J2000) 
  $-$24\arcdeg22\arcmin43\farcs0 \citep{mai04}.
\label{lp}}
\end{figure}

\section{Observation and Data Reduction}

\subsection{Thermal Near-infrared Imaging}

The infrared imaging at $L'$ (3.8~$\mu$m) and $M'$ (4.7~$\mu$m) was
carried out at the {\it VLT} UT4 on 2003 June 11 with the adaptive
optics imager NACO \citep{rou00,len03,har03}. Herschel~36 ($V =$
9.1~mag) served as a wavefront reference source for the visible
wavefront sensor in the adaptive optics system. A short exposure of
180 ms was repeated 27 times in the $L'$ imaging at each of the 9
positions of the telescope dithering. Imaging at $M'$ was performed
the same way, but with a shorter integration time of 56~ms repeated 89
times. The total on-source integration time is 58~s and 60~s at $L'$
and $M'$, respectively. The observing log is presented in
Table~\ref{tb1} including imaging with other filters and additional
spectroscopy.

The imaging data were reduced in the standard manner. After
sky-subtraction and flat-fielding, images were registered referring to
the position of Herschel~36. The size of the isoplanatic patch was
measured using more than 40 stars inside the entire field of view of
NACO (27\arcsec$\times$27\arcsec). The measurements have been done at
2.2~$\mu$m, since the isoplanatic patch becomes smaller with the
wavelength. The point spread function (PSF) is found elongated only at
the edge of the field of view, and no significant degradation of PSF
is recognized within the field of view relevant to the following
discussion shown in Figure \ref{lp}. The full width at half maximum
(FWHM) of point sources are 0\farcs11 at $L'$ and 0\farcs13 at $M'$ in
the fully reduced images.

\vspace{2mm}
\figurenum{2}
\begin{figure}[b]
\includegraphics[bb=15 5 385 793,angle=0,width=0.21\textwidth]{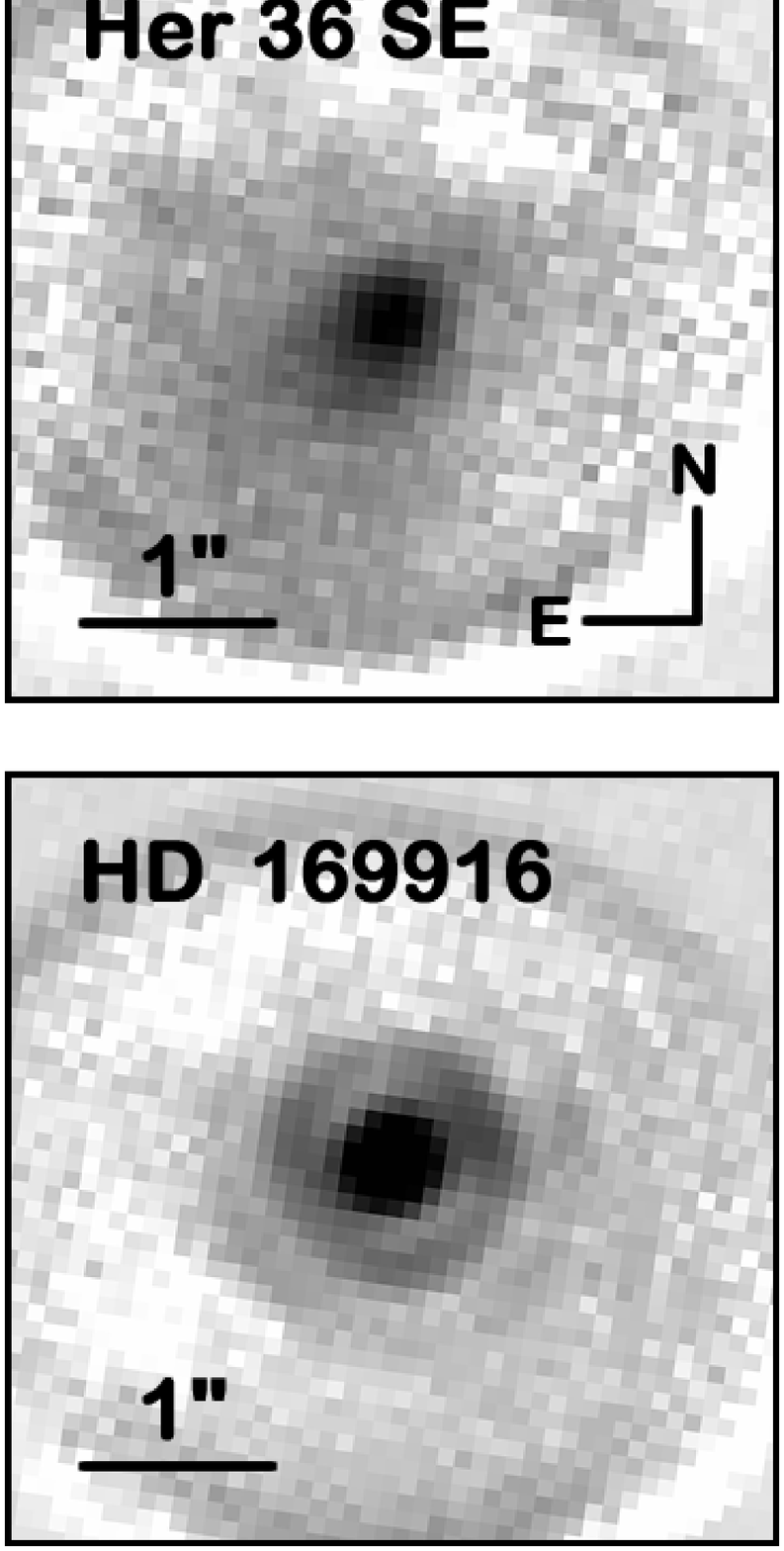}
\includegraphics[bb=113 56 485 750, angle=0,width=0.25\textwidth]{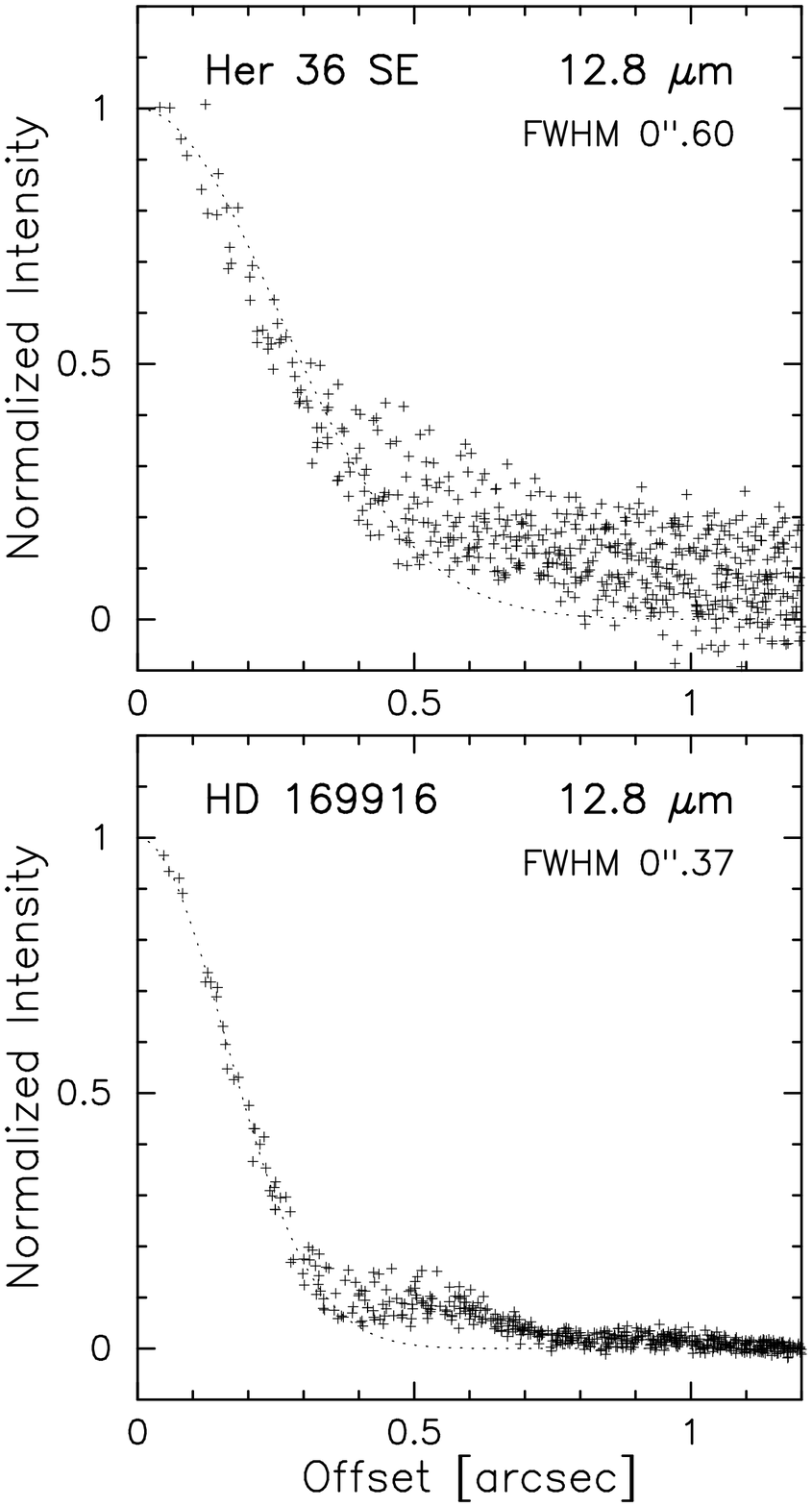}
\caption{Left: MIDI source-acquisition images at 12.8~$\mu$m for
  Her~36~SE (top) and the flux calibration star HD~169916
  (bottom). Herschel~36 is no longer visible at this wavelength. While
  HD~169916 has multiple diffraction rings around, Her~36~SE is
  extended with no trace of a point source. Right: The radial profile
  of Her~36~SE and HD~169916. The size of the emitting region at
  Her~36~SE is 850~AU in diameter, after deconvolution with the
  spatial profile of HD~169916. 
\label{psf}}
\end{figure}

\begin{deluxetable*}{llccclll}[b]
\tablecaption{\label{tb1}}
\tablewidth{\textwidth}
\tablehead{
\colhead{UT Date}            &
\colhead{Filter/Band}        &
\colhead{$\lambda$ [$\mu$m]} &
\colhead{$\Delta \lambda$\tablenotemark{a} [$\mu$m]} &
\colhead{Telescope         } &
\colhead{Instrument        } &
\colhead{Observation       } &
\colhead{Spectral Resolution} 
}
\startdata

2003 Jun 11  & $L'$             & 3.80 & 0.62   & {\it VLT} & NACO & Imaging & \\
2003 Jun 11  & $M'$             & 4.78 & 0.59   & {\it VLT} & NACO & Imaging & \\
2003 Jun 11  & Br$,\gamma$      & 2.17 & 0.023  & {\it VLT} & NACO & Imaging & \\
2003 Jun 11  & IB~218           & 2.18 & 0.060  & {\it VLT} & NACO & Imaging & \\
2004 Jun 3   & N8.7             & 8.64 & 1.55   & {\it VLT} & MIDI & Imaging & \\
2004 Jun 3   & [\ion{Ne}{2}]    & 12.8 & 0.39   & {\it VLT} & MIDI & Imaging & \\
2004 Jul 29  & Br$\,\alpha$     & 4.05 &        & Subaru    & IRCS/AO & Spectroscopy & $R = 10,000$ \\
\enddata
\tablenotetext{a}{In FWHM.}
\end{deluxetable*}

The absolute flux calibration was tied to the photometry of
Herschel~36 as given in the literature. The zero point magnitudes were
calculated to be consistent with the photometric magnitudes $L' =$
6.3~mag and $M' =$ 3.8~mag in \citet{woo90} and \citet{woo73},
respectively. The images were convolved with Gaussian filter at the
zero-point calculation to match the spatial resolution in the previous
observations. The photometry of Herschel~36 was performed inside a
small aperture of 0\farcs2 to avoid confusion with Her~36~SE. An
aperture correction was applied by using the PSF sampled at Her~36~B
(see Figure~\ref{lp}).  The photometry of Her~36~SE was then performed
in a circular aperture of 1\farcs3, after the contribution of
Herschel~36 was removed by subtracting the scaled PSF. The primary
error source in the photometry is the spatial fluctuation of the
background emission in the immediate vicinity of Herschel~36. We
restored the flux compensation function with varying outer bounds from
0\farcs6 to 1\farcs4, and found that the amount of aperture correction
does not differ more than 15~\%. The sky level sampled at different
locations at 0\farcs9 to 2\farcs6 around the object does not change
the net photometry by more than 14~\%. We therefore quote 0.2~mag as
the photometric accuracy, although the formal error is much smaller
($<$0.05~mag). The results are presented in Table~\ref{tb2} with other
photometry obtained in this paper. The color composite image of $L'$,
$M'$ and the narrow-band image at 2.17~$\mu$m described in the next
section is shown in Figure~\ref{lp}.

\subsection{Narrow-band Imaging at Br$\,\gamma$}

Narrow-band Br$\,\gamma$ images (2.17~$\mu$m) were obtained during the
same night together with continuum images at 2.18~$\mu$m. Short
exposures of 1~s were repeated 20 times to minimize the saturation of
bright stars. The total integration time on source is 100~s. The sky
at 30\arcsec~distance from the field was recorded for background
subtraction. The data were processed in the same way as it was the
case for broadband imaging. Aperture photometry was performed for
Her~36~SE in Br$\,\gamma$ and the continuum at 2.18~$\mu$m.  Since
Herschel~36 is saturated, Her~36~B was used to establish the correct
flux scale of the images ($K = 9.4$~mag; KS1 in Woodward et al. 1990,
18006nr766 in Bik 2004). The PSF sampled from Her~36~B was scaled and
subtracted from Herschel~36 to isolate the extended emission of
Her~36~SE. The total pixel counts were summed up inside a circular
aperture of 1\farcs3 centered on Her~36~SE. Note that the Br$\,\gamma$
photometry presented in Table~\ref{tb2} is before the continuum
subtraction. The accuracy of the photometry is $\sim$0.1~mag for the
images with the underlying continuum emission.  The continuum image
was scaled and subtracted so that the pixel counts of blue stars (with
respect to their infrared colors) around Herschel~36 are equally
canceled in the line-emission image.

\subsection{Mid-infrared Imaging at 8.7~$\mu$m and 12.8~$\mu$m}

Mid-infrared images in the $N1$ filter (8.7~$\mu$m) and in the
[\ion{Ne}{2}] filter (12.8~$\mu$m) were obtained on 2004 June 3 at the
{\it VLT} with MIDI \citep{lei03}. The instrument is an
interferometer/spectrometer, but was used as mid-infrared camera in
the present observation. The tip-tilt corrector STRAP was used to
stabilize the images. These thermal infrared data were recorded by
using a chopping throw of 10\arcsec. The total on-source integration
is 80~s at 8.7~$\mu$m and 375~s at 12.8~$\mu$m. The data reduction was
carried out using the pipeline provided by the MIDI
consortium\footnote{http://www.mpia-hd.mpg.de/MIDISOFT/}. The spatial
resolution of the final image is nearly diffraction limited
(Fig.~\ref{psf}). 

\begin{figure}[tbh]
\figurenum{3} 
\includegraphics[angle=-90,width=0.42\textwidth]{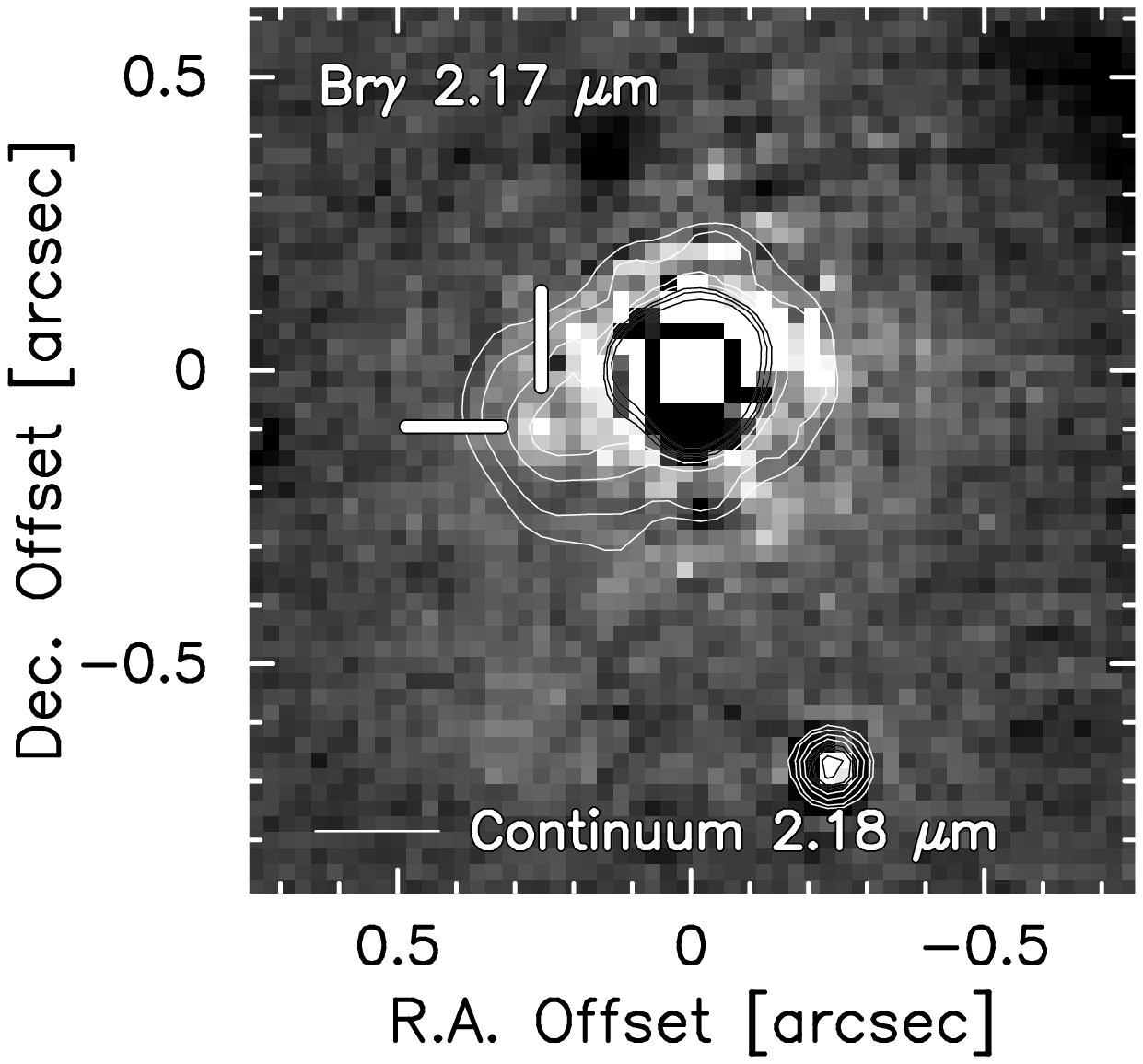}
\includegraphics[angle=-90,width=0.42\textwidth]{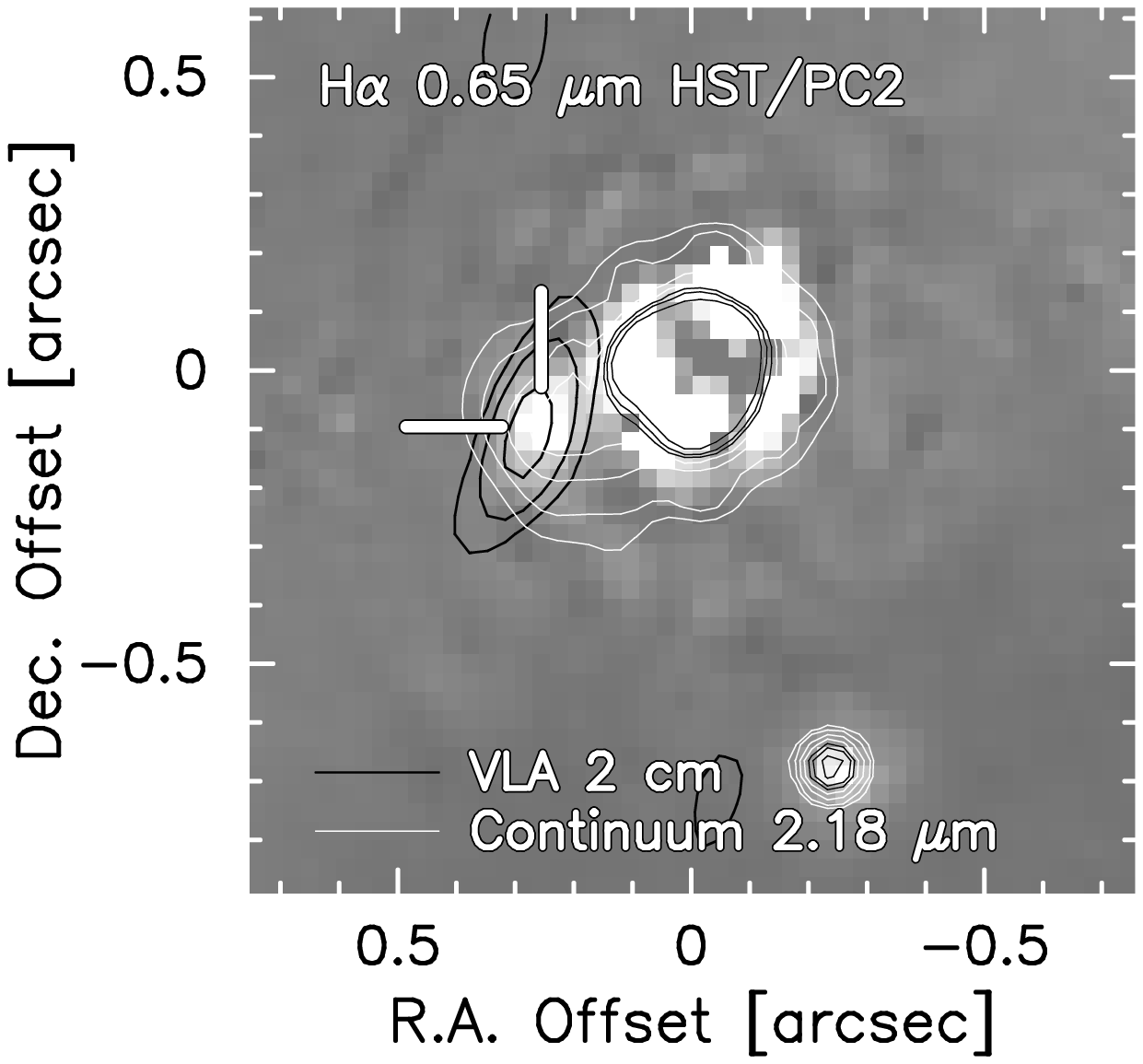}
\caption{(a) Continuum-subtracted Br$\,\gamma$ image of Her~36~SE.
  The location of the point source is marked by ticks. The gray
  contour is from the narrow-band continuum image at 2.18~$\mu$m
  presented darker at around Herschel~36 for clarity. No point source
  is detected at the continuum wavelength. (b) The H$\,\alpha$ image
  retrieved from the {\it HST} archive. The PSF generated by {\it
  TinyTim} is subtracted. The contour in black is radio continuum
  emission at 2~cm obtained with the {\it VLA} (see Stecklum et
  al. 1998 for observational detail). The unresolved sources at
  Br$\,\gamma$, H$\,\alpha$, and radio wavelengths are all in
  positional agreement. The coordinate offsets are with respect to
  Herschel~36. 
\label{ha}}
\end{figure}

The emission from Her~36~SE is clearly extended both in $N1$ and
[\ion{Ne}{2}]. Herschel~36 is no longer visible at mid-infrared
wavelengths, as is expected from its photospheric spectral energy
distribution (SED). It is clear that the peculiar mid-infrared excess
toward Herschel~36 \citep{dyc77} is not from the O-star itself, but is
almost entirely attributed to Her~36~SE. The flux calibration was
performed with respect to the photometric standard HD~169916 for which
the absolute flux density was taken from \citet{coh99}. The photometry
was performed inside a 1\farcs8 aperture centered on Her~36~SE. The
size of the aperture is slightly larger than that is used in the
shorter wavelengths. The smaller aperture at the thermal near-infrared
is because there seem to be two overlapping emission contributions at
Her~36~SE; the compact dusty cloud at Her~36~SE itself, and the
filamentary emission more connected to the diffuse emission at
2\arcsec~southeast of Herschel~36. Since we will discuss an internal
source inside Her~36~SE below, the diffuse emission should be excluded
not to overestimate its luminosity. On the other hand, the
mid-infrared images by MIDI do not show any clear hints of multiple
sources, we therefore use a safe oversized aperture not to lose the
mid-infrared flux of Her~36~SE for later discussion of its energetics.

\begin{figure*}[bth]
\figurenum{4}
\includegraphics[angle=-90,width=\textwidth]{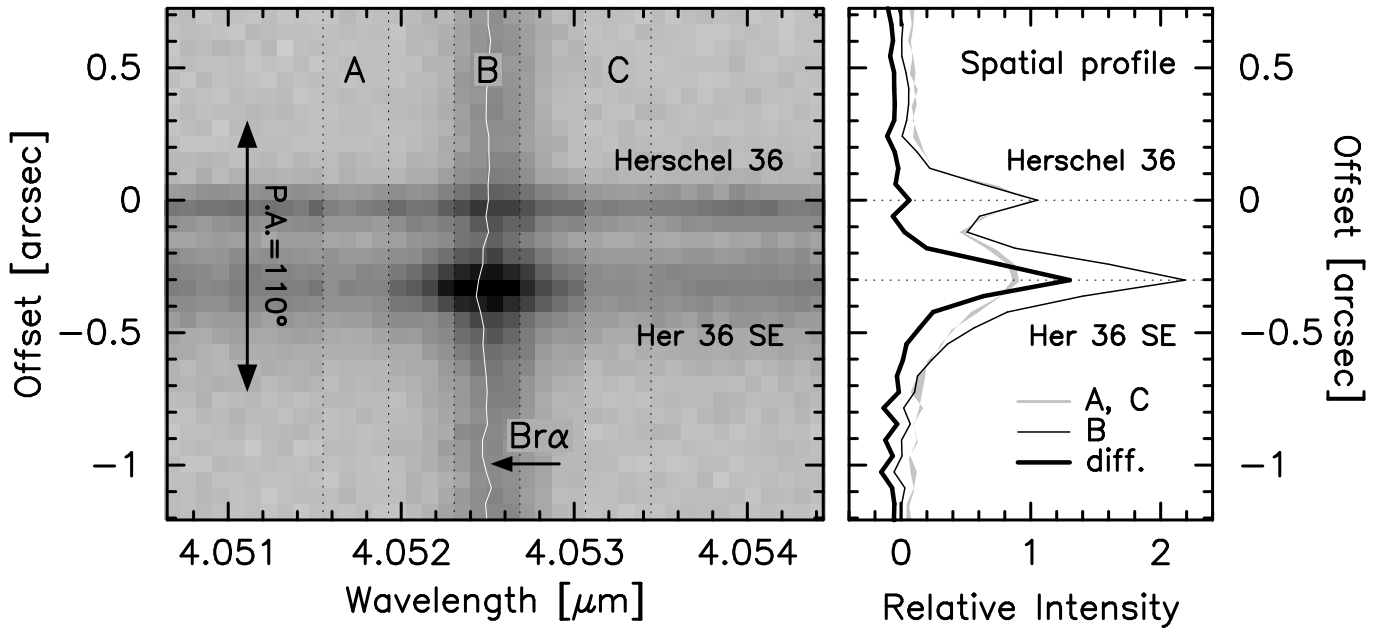}
\caption{Left: A close-up view of the two--dimensional spectrogram
  near the Br$\,\alpha$ emission. The continuum emission at Her~36~SE
  is clearly separated from that of Herschel~36, and apparently
  extended (position A and C). The line emission from the ambient
  nebulosity is seen at Br$\,\alpha$ (position B). The white line
  traces the wavelength centroids of the Br$\,\alpha$ emission along
  the slit. The emission line is slightly ($\sim$2~km~s$^{-1}$)
  blueshifted at Her~36~SE with respect to the ambient nebular
  emission. Right: Cross-cuts of the spectrogram along the slit, at
  the continuum (A and C; gray polygon), and the Br$\,\alpha$ emission
  (B; thin solid line). The contribution of the nebulosity is
  subtracted at the position B. Her~36~SE shows a sharp emission line
  at Br$\,\alpha$ on top of the extended continuum emission. The line
  emission profile, after subtracting the continuum profile, is close
  to that of Herschel~36, which means that the Br$\,\alpha$ emission
  at Her~36~SE is also compact and point-like.
\label{bra}}
\end{figure*}

\subsection{Medium-Resolution Br$\,\alpha$ Spectroscopy}

Supplemental 4-micron spectroscopy ($R = $ 10,000) was performed at
the Subaru Telescope on 29 July 2004 with IRCS \citep{tok98,kob00}.
The slit was oriented along a position angle of 110\degr~counted from
north to east to cover Herschel~36 and SE at the same time. An
adaptive optics system was used to attain higher spatial resolution
\citep{gae02,tak04}. The sky at 2\arcmin~north of Herschel~36 was
observed for background subtraction after each on-source integration.
The spectroscopic flat field was obtained from a halogen lamp exposure
at the end of the night.

One dimensional spectra of Herschel~36 and SE were extracted with the
aperture-extraction package of IRAF\footnote{IRAF is distributed by
the National Optical Astronomy Observatories, which are operated by
the Association of Universities for Research in Astronomy, Inc., under
cooperative agreement with the National Science Foundation.} after the
sky-subtraction, flat-fielding, and the interpolation of the bad
pixels were applied. The wavelength calibration was carried out using
the atmospheric transmission curve modeled by ATRAN \citep{lor92}. The
Br$\,\alpha$ line-emission was calibrated to the photometry of
Herschel~36 at $L'$. First, to correct the continuum slope, the
one-dimensional spectrum of Herschel~36 was divided by the
spectroscopic standard star HR~7121 (B2.5V), and was multiplied by a
blackbody spectrum of the temperature corresponding to the effective
temperature of a B2.5V star ($T_{\rm eff}=$19,000~K; Crowther
2005). The spectrum with correct slope was then scaled so that the
averaged flux density inside the $L'$ bandpath (3.49--4.11~$\mu$m) is
equal to the $L'$-band photometry of Herschel~36. The same conversion
factor was used to calibrate the spectral flux of Her~36~SE reduced in
the same way with Herschel~36. The Br$\,\alpha$ luminosity at
Her~36~SE is found to be $L({\rm Br}~\,\alpha)=$(1.4--1.6)$\times
10^{24}$~W at the distance of 1.8~kpc, after the continuum and the
surrounding diffuse emission is subtracted. The error interval is
given by the difference of the two sequential measurements, although
it is subject to the uncounted uncertainty associated to the possible
vignetting by the narrow slit of 0\farcs3.

\section{Results}

An unresolved source is detected at the location of Her~36~SE in the
continuum-subtracted image at Br$\,\gamma$ (Fig.~\ref{ha}). The
diameter of the point-like source is less than 130~AU from the
diffraction-limited spatial resolution of NACO (0\farcs072 in FWHM at
2.17~$\mu$m). The {\it HST}/PC2 image retrieved from the
ST-ECF\footnote{The Space Telescope European Coordinating Facility
(ST-ECF), jointly operated by ESA and the European Southern
Observatory (ESO), is the European {\it HST} science facility,
supporting the European astronomy community in exploiting the research
opportunities provided by the Hubble Space Telescope.} archive shows a
compact H$\,\alpha$ emission at the same location (Fig.~\ref{ha}). The
H$\,\alpha$ emission is unresolved as well. Considering the plate
scale of PC2 (0\farcs046~pixel$^{-1}$), this finding indicates that
the source is less than 100~AU across. Furthermore, radio
interferometric observations have been carried out with the {\it VLA}
at a wavelength of 2~cm (see Stecklum et al. 1998 for the
observational detail). A compact radio source is found at the same
location as the position of the Br$\,\gamma$ and H$\,\alpha$
emission. The radio source is unresolved with regard to the
synthesized beam size of 0\farcs16, which translates to 290~AU.

Another line of evidence for a point-like source comes from the
spectroscopy. The two-dimensional spectrogram near Br$\,\alpha$ is
shown in Figure~\ref{bra}. Her~36~SE shows distinct line emission in
Br$\,\alpha$, slightly blueshifted from ambient nebular emission by 2~
km~s$^{-1}$, with a spatial profile apparently narrower than the
continuum emission. The sharp spatial profile is comparable to that of
Herschel~36, which corroborates the presence of a point-like source in
the hydrogen line emission at the location of Her~36~SE.

\begin{figure}[b]
\figurenum{5}
\includegraphics[angle=-90,width=0.48\textwidth]{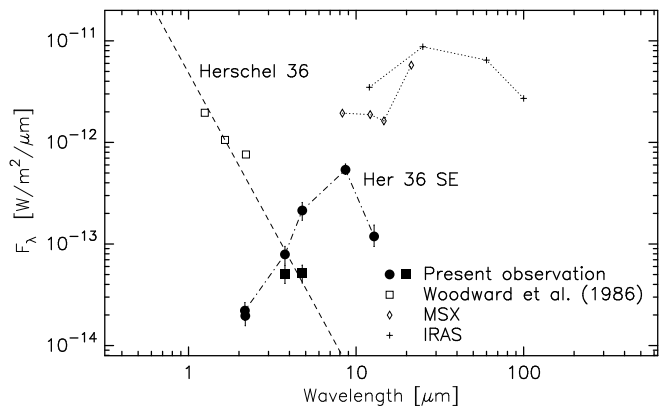}
\caption{The infrared SED of Herschel~36 and SE, plotted by the
  circles and the squares, respectively. The filled symbols are from
  the present observations, and the crosses and open symbols are from
  the literature. Note that this previous photometry has been done
  with much larger apertures. The color temperature of the dust
  emission measures 400~K from 3--13~$\mu$m photometry, which is taken
  as the upper limit of the dust emission temperature of
  Her~36~SE. 
\label{sd}}
\end{figure}

On the other hand, Her~36~SE is clearly extended in the continuum
emission at wavelengths from 2 to 13~$\mu$m. If we use the spatial
profile of HD~169916 as the instrumental PSF, and deconvolve Her~36~SE
by inverting simple square sum, the extent of the emitting source at
Her~36~SE is reduced to 0\farcs47 at 12.8~$\mu$m, which is 850~AU in
diameter at the distance of the object (Fig.~\ref{psf}).  The SED of
Her~36~SE is presented in Figure \ref{sd} to characterize the extended
emission. The color temperature of the continuum source clearly points
to the existence of warm dust at the location of Her~36~SE. No
point--like substructure is found in the broad--band images of
Her~36~SE that could have corresponded to the unresolved line
emission.

\section{Discussion -- Nature of Herschel~36~SE}

Here we first discuss the identity of the line emission source and its
possible ionization mechanism, including external ionization by
Herschel~36, an embedded low- to intermediate-mass star in its active
accretion phase, and an \ion{H}{2} region internally ionized by an
early-type star.

The Br$\,\gamma$ emission is apparently inside the dusty cloud at
Her~36~SE, since it is spatially more confined than the continuum
emission. In addition, there is no hint of rim-ionization detected in
Br$\,\gamma$ emission at the side of Her~36~SE where it faces toward
Herschel~36. We found no solid evidence that Herschel~36 plays a
direct role to externally ionize the unresolved source inside
Her~36~SE.

The radio flux at 2~cm ($F_{\nu} = $1.3~mJy) is probably too high for
an accretion signature of an intermediate-mass star at 1.8~kpc
away. \citet{neu96} have calculated the free-free emission arising
from an accretion shock in dependence of (proto)stellar mass and
accretion rate. However, even with their most extreme setup ($M$ = 10
$M_{\odot}$, $\dot{M}_{\rm acc} = 10^{-4} M_{\odot}$yr$^{-1}$) they
just reach a 3.6 cm flux of roughly 3 mJy for a source 100 pc
away. Extrapolated to $\lambda = 2$~cm (by optimistically assuming
that the ionized gas is completely optically thick with $F_\nu ~\sim
\nu^2$) and scaled to a distance of 1.8 kpc, the expected 2-cm flux
would be just some 30 $\mu$Jy, around 40 times less than the measured
value.

It is therefore inferred that a star with an early spectral type
exists inside Her~36~SE that gives rise to a small \ion{H}{2} region
responsible for the radio emission. The \ion{H}{2} region is
internally ionized, but is kept compact because of the high density of
the enshrouding cloud. If we take the {\it VLA} beam size as the
physical dimension of the \ion{H}{2} region ($r\sim$140~AU), the
number of Lyman continuum photons required to maintain the ionized gas
is 1.6$\times$10$^{45}$ s$^{-1}$. The Lyman continuum photon-rate is
reproducible only by a star earlier than B2 if the star is at the
zero-age main sequence \citep{pan73,cro05}. On the other hand, in
order not to create a parsec-scale \ion{H}{2} region, the hydrogen
density has to be as high as $n_{\rm H} > 10^6$~cm$^{-3}$.

\begin{table*}[tbh]
\begin{center}
\caption{   \label{tb2}}
\begin{tabular}{lccc}
\tableline\tableline
               & 
Herschel~36    &
Her~36~SE \tablenotemark{a} \\


{Filter           }&
{$F_\lambda$ \tablenotemark{b} [$10^{-14}$ W m$^{-2}$ $\mu$m$^{-1}$]} &
{$F_\lambda$ \tablenotemark{b} [$10^{-14}$ W m$^{-2}$ $\mu$m$^{-1}$]}  &
{PSF in FWHM} \\

\tableline
Br$\,\gamma$     & \nodata             & 2.2 \tablenotemark{c}  & 0\farcs072\\
IB~218           & \nodata             & 2.0 & 0\farcs071\\

$L' $            & 5.1                 & 7.9 & 0\farcs11\\
$M' $            & 5.2                 & 21  & 0\farcs13\\
N8.7             &\nodata              & 54  & 0\farcs35\\
\ion{Ne}{2}      &\nodata              & 12  & 0\farcs37\\

\tableline
\end{tabular}
\tablenotetext{a}{The photometric apertures are 1\farcs3
in the mid-infrared bands (N8.7 and \ion{Ne}{2}), and 1\farcs8 
in all the others.}
\tablenotetext{b}{The photometric errors are typically 10\%
at 2.2~$\mu$m, and 20\% at all the other filters.}

\tablenotetext{c}{Continuum is not subtracted.}

\end{center}
\end{table*}

We use the $K$-band extinction toward the hypothetical early-type star
to estimate whether the gas density is sufficiently high to confine
the \ion{H}{2} region. A B2 star at the zero-age main-sequence should
have a relative $K$-band brightness of 9.9~mag at the distance of
Herschel~36 without any attenuation ($(V-K)_0=-0.9$~mag from Ducati et
al. 2001). The sensitivity of our observation at 2~$\mu$m is 15.7~mag
for a 3 sigma detection at the location of Her~36~SE. With
non-detection of any continuum point source at this wavelength, the
dust extinction must be larger than 6~mag at 2~$\mu$m, which
translates to $A_V>$ 60~mag after correcting $A_V\approx$ 5~mag for
the foreground extinction toward the Herschel~36 region \citep{ste98}.
The visible extinction can be related to a hydrogen column density
\citep[e.g., ][]{mat90,ryt96}. Provided that the dusty core of
Her~36~SE is spherical, of constant density, and 850~AU across as is
measured in the 12.8~$\mu$m image; the mass in the obscuration is
$M_{\rm SE} \ge 1.7 \times 10^{-2} M_\odot$ with $n_{\rm H} \ge 1.8
\times 10^7$~cm$^{-3}$. Thus, the gas density should be high enough to
keep the \ion{H}{2} region spatially unresolved.

The Lyman photon rate derived from Br$\,\alpha$ spectroscopy is also
consistent with that of an early B type star. The ionizing flux in an
\ion{H}{2} region is obtained from Br$\,\alpha$ line flux by scaling
the photon number-count proportionally to the recombination
coefficients, ${\int_{\nu_0} L_\nu / h\nu~d\nu} \approx {\alpha_{B}} /
{\alpha^{\rm eff}_{\rm Br\,\alpha}} \cdot L({\rm
Br\,\alpha})/{h\nu_{\rm Br\,\alpha}} $. If we take $\alpha^{\rm
eff}_{\rm Br\,\alpha} = 1.085 \times 10^{-14}$cm$^{3}$~s$^{-1}$ and
$\alpha_{B} = 2.658\times 10^{-13}$cm$^{3}$~s$^{-1}$ from
\citet{sto95} for the Case~B of $T_{\rm e} = 10^4$~K and $N_{\rm e} =
10^7~$cm$^{-3}$, the Lyman continuum rate turns out (6.9--8.0)
$\times$10$^{44}$ s$^{-1}$, which is reproducible by a B2.5 dwarf
\citep{cro05}. The correction of the foreground extinction needs
caution, since the dust obscuration is increasingly transparent in the
longer wavelength \citep{rie85}; but if we use $A_K = 6$~mag as a face
value, the intrinsic Lyman continuum rate is (4.1--4.8)
$\times$10$^{45}$ s$^{-1}$, which is still consistent with the
ionizing photon rate of a B1--B1.5 dwarf.

The infrared luminosity of Her~36~SE is consistent both with an
internal B2 star, and also with external heating by Herschel~36. The
infrared luminosity from 2 to 40~$\mu$m is calculated from the NACO
and MIDI photometry with the flux density at the longer wavelength
extrapolated as $ F(\lambda) = \kappa(\lambda)~M_d ~d^{-2}~B(T_d,
\lambda)$; where $d$ is the distance to the object, and
$\kappa(\lambda)$ is the computed mass absorption coefficient for the
grains without ice mantles coagulated in the protostellar cores
\citep{oss94}. The total infrared luminosity is $L_{\rm IR} =
400~L_\odot$ at an assumed distance of 1.8~kpc insensitive to the gas
density of the core $n=10^6$ cm$^{-3}$ to $10^8$~cm$^{-3}$. It is
therefore well reproducible either by the luminosity of a B2 star at
the zero-age main-sequence ($L_\ast = 3 \times 10^3~L_\odot$), or by
Herschel~36 ($L_\ast = 10^5~L_\odot$) while the solid angle subtended
by Her~36~SE is of the order of unity at the location of
Herschel~36. The dust emitting temperature is 400~K which should be
taken as the upper limit, for the lack of additional photometry at the
longer wavelengths.

We conclude by comparing Her~36~SE with two similar cases reported to
date in which bright mid-infrared sources are found in the immediate
(projected) vicinity of O-type stars. The infrared source SC3 has been
found at 1\farcs8 (810~AU) west of $\theta^1$~Ori~C, the primary
illumination source of the Orion Trapezium Cluster \citep{hay94}. SC3
is spatially resolved, measuring 1\farcs5 across, however, despite the
apparent proximity to $\theta^1$~Ori~C (O5.5V), its appearance is
barely distorted, almost with a perfect circular symmetry. It is thus
proposed that SC3 is a proplyd seen face-on, located deep behind
$\theta^1$~Ori~C with the physical separation much larger than the
apparent projection \citep{rob02}. SC3 is visible in the optical
\citep[e.g., ][]{smi05}, and at near-infrared wavelengths
\citep{mcc94}, which also lends support to its proplyd nature. The
infrared appearance of SC3 is in strong contrast to Her~36~SE. We may
use the highly distorted dust emission from Her~36~SE as
circumstantial evidence that the source is actually under the
influence of Herschel~36, and that the physical distance to the O star
is not significantly larger than it appears.

On the other hand, $\sigma$~Ori~IRS~1, found next to $\sigma$~Orionis,
shares a similar morphology with Her~36~SE. It is a compact infrared
source at 1200~AU away from the O9.5V star, and features a fan-shaped
emission with $\sigma$~Orionis at the apex \citep{loo03}, exactly as
Herschel~36 is to SE. The mid-infrared spectrum of $\sigma$~Ori~IRS~1
shows the partly crystalline silicate in emission. The presence of
processed silicates suggests significant grain growth which is
naturally present in a circumstellar disk. A proplyd is therefore
again the most probable cause of $\sigma$~Ori~IRS~1, especially
because a central star has been detected recently in the $K$-band
continuum emission (B. Stecklum, private communication).

The star-forming region around Herschel~36 has many features in common
with the Orion Nebula Cluster. The local concentration of massive
stars, like Her~36~B, together forms a Trapezium-like cluster around
Herschel~36. The presence of known proplyd nearby at G5.97$-$1.17
\citep{ste98} underscores the physical similarity as well. The
mid-infrared color of Her~36~SE, and the close vicinity to an O-type
star with a distorted morphology suggestive of radiative influence of
it all point toward Her~36~SE is also a proplyd with a low-mass star
at its center, as is the case for SC3 at $\theta^1$~Ori~C and
$\sigma$~Ori~IRS~1. However, in addition to the radio luminosity
hardly accounted for by a low-mass star, and no ionized-front outside
the dusty cloud; a proplyd cannot explain the lack of a point source
to be detected at the continuum wavelengths that comes from the
photospheric emission of the star. In the case of no central star
inside, another possibility would be that Her~36~SE is an evaporating
gaseous globule, a failed proplyd without an internal star in
formation. These starless cores have been detected in numbers toward
M~16 \citep{mcc02}. However, a hypothetical starless globule conflicts
with the presence of the unresolved Br$\,\gamma$ emission apparently
inside of Her~36~SE. We therefore propose that Her~36~SE harbors a
relatively massive star of early B-type producing a squeezed
\ion{H}{2} region inside the dusty cloud \citep{ket03}, but the star
itself is completely obscured. The distortion of dust emission as well
as the diffuse emission downstream of Her~36~SE, indicates the close
physical interaction of Herschel~36 and SE. Herschel~36~SE, now in the
process of being blown away, is a showcase for the violent impact of
the dominant O-star in a cluster on another early-type star nearby.

\acknowledgments

{We thank all the staff and crew of the {\it VLT} and Subaru for their
valuable assistance in obtaining the data, and Thorsten Ratzka, Elena
Puga and Wolfgang Brandner for their indispensable help in reducing
data. We appreciate the anonymous referee for many critical comments
that are necessary to improve the paper. M.G. is supported by Japan
Society for the Promotion of Science fellowship.}

\end{document}